\documentstyle[preprint,aps,prl]{revtex}

\newcommand{\beq}{\begin{equation}}
\newcommand{\eeq}{\end{equation}}
\newcommand{\beqa}{\begin{eqnarray}}
\newcommand{\eeqa}{\end{eqnarray}}

\newcommand{\bi}{\bibitem}

\newcommand{\rd}{\rho_{\rm d}}

\begin{document}

\draft
\title{
Metal-insulator transition in 2D:\\
Anderson localization by temperature-dependent
disorder?}
\author{B.\ L.\ Altshuler$^{1,2}$, D.\ L.\ Maslov$^{3,}\footnote{Corresponding
 author.\\
E-mail: maslov@phys.ufl.edu
Tel. (352) 392-0513.
Fax: (352) 392-0524}$,
 and V.\ M.\ Pudalov$^{4}$}
\address{
$^{1)}$NEC Research Institute, 4 Independence Way, Princeton, NJ 08540\\
$^{2)}$Physics Department, Princeton University, Princeton, NJ 08544\\
$^{3)}$Department of Physics, University of Florida\\
P.\ O.\ Box 118440, Gainesville, Florida 32611-8440\\
$^{4)}$P. N. Lebedev Physics Institute, Russian Academy of Sciences\\
Leninsky Prospect 53, Moscow 117924, Russia
}
\maketitle
\begin{abstract}
A generalization of the single-parameter scaling 
theory of localization is proposed for the case
when the random potential depends 
on temperature. The scaling equation describing
the behavior of the resistance is derived. It is shown
that the competition between the metallic-like 
temperature dependence of the Drude resistivity
and localization leads to
a maximum (minimum) at higher (lower) temperatures.
An illustration of a metal-insulator transition 
in the model of charged traps whose concentration
depends on temperature is presented.
\end{abstract}
\bigskip
\pacs{PACS numbers: 71.30.+h, 72.15 Rn, 73.40.Qv}
Until recently, the one-parameter scaling theory of localization (STL)
\cite{gang4},
supplemented by the perturbative treatment of interactions between
electrons, had enjoyed an overwhelming agreement with the experiment. This
agreement strengthened a common belief that the STL's description of weak
and strong localization regimes, as well as of 
crossover between the two, is qualitatively correct even in the
presence of interactions. 
Recent observations of an unexpected metallic
behavior in 2D
electron- and hole-gas systems, starting from the pioneering work
by Kravchenko et al. 
\cite{kravchenko}, have raised
concerns about the general validity of the STL.
It has also been suggested that the metallic-like temperature dependence of the
resistance $\rho$ is a signature of
a novel and perhaps non-Fermi-liquid state, which either results 
from the interplay of strong Coulomb interactions and disorder
or is a non-trivial superconducting
state \cite{nfl}.
We believe though that a number of experimental results indicates strongly
that the metallic state is quite conventional and
manifestly Fermi-liquid--like, and that the $\rho(T)$-dependence
is not determined by the interactions.
These results include:\\
i) well-pronounced Shubnikov-de Haas oscillations in the
same systems that demonstrate a strong $\rho(T)$-dependence
at higher temperatures \cite{vmp_shdh};\\
ii)negative magnetoresistance of the weak-localization (WL) type at low
temperatures \cite{vmp_nmr};\\
iii)~almost absent $T$-dependence of the low-field Hall
resistance $\rho_{xy}$ 
(as compared to that of $\rho_{xx}$) \cite
{hall}. (If  the correction to $\rho$ were due to
electron-electron interactions in a disordered system, the following
relation should have been satisfied: 
$\delta\rho_{xy}/\rho_{xy}=2\delta\rho_{xx}/\rho_{xx}$ \cite{aa});\\
iv)~an insulating upturn of the resistivity at low temperatures and
high densities \cite{vmpsept98};\\
v) ~persistence of the metallic-like $\rho(T)$-dependence up to
densities of about $30$ times larger than the density at the transition and,
correspondingly, down to resistivities of about $100$ times smaller than the
resistivity at the transition \cite{vmp_drude}. In this region, 
quantum interference of interacting electrons
is already amenable to
the perturbative treatment
\cite{aa} and is known to result only in small (of the order of few $\%$s) 
corrections to the resistance, whereas the observed
variation of $\rho$ still constitutes $200-300\%$.
This suggests that the $\rho(T)$-dependence in the
metallic phase is not due to quantum effects of interference
and interactions.

With this in mind,  we are going to generalize the STL for the case
when the disorder potential has two components: a $T$-independent
and $T$-dependent one. The latter component may result from, {\it e.g.},
charged defects (traps) whose concentration changes with $T$\cite{am1}
or from the $T$-dependence of the screening radius \cite{dassarma}.
Correspondingly,
the Drude (classical) resistance of such a system
is 
\begin{equation}
\rho_{{\rm d}}(T)=\rho_1+\rho_0(T),  \label{drude_gen}
\end{equation}
where $\rho_1$ is the $T$-independent residual resistance, whereas
$\rho_0$, resulting from scattering at the $T$-dependent component
of disorder, is metallic-like ($d\rho_0/dT>0$).
The goal of this work is: given
that $\rho_{{\rm d}}(T)$ is of the form (\ref{drude_gen}), to account for
localization effects and to describe the crossover from the metallic to the
insulating phase.

Let us choose $T$ to be low enough
so that the phase-breaking length $L_{\varphi }$ is much
larger than the system size $L$.
Once $T$, and thus the random potential is fixed, we perform the
conventional scaling of $\rho $ with $L$, which 
leads to an RG equation \cite{gang4}: 
\begin{equation}
\partial \ln \rho (L,T)/\partial \ln L=-\beta (\rho ),  \label{opst}
\end{equation}
where $\beta (\rho )$ is the same 
scaling function  which appears in the STL. The only difference
between Eq.~(\ref{opst}) and the STL
is that now $\rho $ depends on both $L$ and $T$, the latter entering
the equation as a parameter, via the $T $-dependence of disorder. 
Integrating Eq.~(\ref{opst}), we get 
\begin{equation}
\ln L/\ell =-\int_{\rho _{{\rm d}}}^{\rho }d\rho ^{\prime }/\rho ^{\prime
}\beta (\rho ^{\prime }),  \label{intopst}
\end{equation}
where $\rho _{d}\equiv \rho (L=\ell (T),T)$ 
and $\ell (T)$ is the
($T$-dependent) elastic mean free path. The $L$-dependence of $\rho $%
, given (implicitly) by Eq.~(\ref{intopst}), crosses over to the $T$%
-dependence when $L$ becomes comparable to $L_{\varphi }$. To account for
this crossover, we substitute $\ln L/\ell \rightarrow \ln L_{\varphi }/\ell $
in Eq.~(\ref{intopst}) and differentiate both side of this equation with
respect to $T$, taking into account the $T$-dependences of both $\rho _{d}$
and $L_{\varphi }$. When doing this, we assume that $\rho _{d}$ depends on $%
T $ only via $\ell (T)$, {\it i.e.}, $\rho _{d}\propto \ell (T)^{-1}$.
The resulting
equation reads: 
\begin{equation}
\frac{1}{\beta (\rho )}\frac{d\ln \rho }{d\ln T}=-\frac{d\ln L_{\varphi }}{%
d\ln T}+\left[ \frac{1}{\beta (\rho _{{\rm d}})}-1\right] \frac{d\ln \rho _{%
{\rm d}}}{d\ln T}.
\end{equation}
The phase breaking time $\tau _{\varphi }$ usually diverges as a negative
power of $T$ at $T\rightarrow 0$. In simplest cases, e.g., when dephasing is
due to phonons, $\tau _{\varphi }$ does not depend on disorder. In other
cases, $\tau _{\varphi }$ may depend on disorder itself. We will consider a general case, when 
$\tau _{\varphi }\propto T^{-p}\rho _{{}}^{1-2\gamma }$,
$p$ and $\gamma$ being some constants. Note that 
$\tau _{\varphi }$ depends on the observable, rather than Drude,
resistance. $L_{\varphi }$ is determined by $\tau _{\varphi }$ and
by the observable diffusion constant of electrons
 $D=\propto \rho ^{-1}$:
\begin{equation}
L_{\varphi }=\sqrt{D\tau _{\varphi }}\propto \sqrt{\tau _{\varphi }/\rho }%
\propto \rho _{{}}^{-\gamma }T^{-p/2}.  \label{lphi}
\end{equation}
With the dependences of $L_{\varphi }$ on $T$ and $\rho $ taken into
account, the RG-equation reduces to 
\begin{equation}
\left[ \frac{1}{\beta (\rho )}-\gamma \right] \frac{d\ln \rho }{d\ln T}=%
\frac{p}{2}+\left[ \frac{1}{\beta (\rho _{{\rm d}})}-1\right] \frac{d\ln
\rho _{{\rm d}}}{d\ln T}.  \label{rgrho}
\end{equation}
Eq.~(\ref{rgrho}) is supplemented by the boundary condition
$\rho (T_{0})=\rho _{{\rm d}}(T_{0})$,
where $T_{0}$ is a cut-off temperature at which localization effects become
negligible, {\it i.e.}, 
$L_{\varphi}(T_0)=\ell(T_0)$.
Eq.~(\ref{rgrho}) is the main technical result of this paper which allows
one to find the observable $\rho(T)$-dependence, arising both
from the phase-breaking processes [the first term on the RHS] and the $T$%
-dependence of the classical resistance $\rho _{{\rm d}}$ [the second term]. 
The $T$-dependence of the random potential implies that, in fact, disorder
is {\it dynamic}.
For our approach to be valid, the time scale of disorder variation $\tau
_{dis}$ should be larger than $\tau _{\varphi }$ but 
smaller than the measurement time.
In what follows, we will focus on the 2D case, for which the 
$\beta$-function can be approximated with a reasonable accuracy by
$\beta (\rho )=-\ln (1+a\rho )$,
where $a=2/\pi$ and $\rho$ is a resistance per square measured in units
of $h/e^2$.

A number of conclusions about the behavior of $\rho $ can be extracted just
from the general form of Eq.~(\ref{rgrho}):
i)~a metallic (insulating)-like $T$-dependence of $\rho _{{\rm d}}$
impedes (facilitates) Anderson localization;
ii)~the system remains metallic, i.e., $d\ln \rho /d\ln T>0$ at $%
T\rightarrow 0$, provided that $\rho _{d}$ vanishes with $T$ rapidly enough,
namely, $\rho _{{\rm d}}<1/\ln 1/T$;
iii)~if $\rho _{d}(T=0)\equiv \rho _{d0}\neq 0$, then the system
will eventually become an insulator at low enough $T$. This happens, however,
only at {\it exponentially} low temperatures [$\propto \exp (-\rho
_{d0}^{-1})]$.

When both $\rho$ and $_{\text{ }}\rho _{d}$ are small, Eq.~(\ref{rgrho}) gives 
$\rho^{-1}(T)-\rho^{-1}{{\rm d}}(T)=-a\ln \left[ L_{\varphi
}(T)/\ell (T)\right]$, which is simply a  WL result.

It is instructive to analyze the case of a
power-law, metallic-like $T$-dependence of $\rho _{{\rm d}}$: 
\begin{equation}
\rho _{{\rm d}}(T)=\rho _{0}(T)={\bar\rho}\left( T/T_{0}\right) ^{q},\;q>0.
\label{power}
\end{equation}
In this case, Eq.~(\ref{rgrho}) takes the form 
\begin{equation}
\left[ \frac{1}{\beta (\rho )}-\gamma \right] \frac{d\ln \rho }{d\ln T}=%
\frac{p}{2}+q\left( \frac{1}{\beta (\rho _{d})}-1\right).  \label{rgpower}
\end{equation}
As $T\to 0$, $\rho _{{\rm d}}\to 0$ and $\beta(\rd)\to 0$.
The second term on the RHS of Eq.~(\ref{rgpower}) 
diverges and a metallic-like dependence (\ref{power})
prevails. At {\it higher} temperatures,
however, when $\rho _{{\rm d}}$ is still large enough,
the localization may be
effective. If $p$ and ${\bar\rho}$ are large enough, $\rho $
first {\it increases} with decreasing $T$, goes through a maximum and then
decreases, approaching the classical $T$-dependence (\ref{power}) for $T\to 0$ [cf. Fig.~%
\ref{fig:powera}, top solid curve].
 The maximum in $\rho$ exists if 
\begin{equation}
{\rm i)}~p>2q\;{\rm and}\;{\rm ii)}~{\bar\rho}>\rho_c=\frac{1}{a}\left[\exp
\left( \frac{1}{(p/2q)-1}\right) -1\right].  \label{crit}
\end{equation}
As ${\bar\rho}$ approaches $\rho_c$ from above, the maximum becomes more
shallow (Fig. ~\ref{fig:powera}, medium solid curve) and it disappears for ${%
\bar\rho}< \rho_c$ (Fig. ~\ref{fig:powera}, bottom solid curve).
).

Even if the resistance is metallic-like at higher temperatures, the
$T$-independent part of disorder leads eventually to localization.
Therefore, $\rho$ should have a minimum at low temperatures.
If the residual resistance $\rho_1\ll 1$,
this minimum occurs already in the WL regime. The temperature of the minimum
is to be determined from the following equation
\begin{equation}
T_{\text{min}}=\frac{pa}{2}\frac{\rho _{1}^{2}}{d\rho _{0}/dT|_{T=T_{\text{min}}}}.
\label{tmin}
\end{equation}
In all experiments on the metal-insulator transition (MIT) in 2D, both $\rho(T)$ and $d\rho(T)/dT$
decrease as the carrier density increases, {\it i.e.}, as the
system becomes more metallic.  Eq.~(\ref{tmin})
tells then that the minimum is shifting to {\it higher} temperatures
as one is getting {\it deeper} into the metallic phase.
This behavior is illustrated in Fig.~\ref{fig:minpower} for $\rho _{0}(T)$ given by Eq.~(\ref{power}).

We now turn to a particular model for $\rd(T)$ \cite{am1},
which describes scattering of 2D electrons at charged traps located in the
oxide layer of thickness $d$ and dielectric constant $\epsilon _{ox}$. The
energy level of a trap is shifted by the electrostatic field between the
gate and the 2DEG as well as by that between the trap and its image charge
in the 2DEG. The competition between these two forces leads to a sharp peak
in the spatial distribution of {\it charged} traps. 
The resistance is obtained by integrating the scattering cross-section due
to a single trap over the spatial distribution of the traps.
In what follows, we assume that $T\ll E_{F}$, and that the chemical
potential $\mu$ of the 2D gas is determined entirely by the gate voltage $V_g$,
and is
therefore $T$-independent. The $\rd(T)$-dependence results
then only from the  thermal population of charged traps.
Let $\mu$ coincide with the trap level 
at $V_{g}=V_{g}^{c}$. We call this
point \lq\lq critical\rq\rq\/, although what appears to be a critical point of
the metal-insulator transition may be shifted from $V_{g}^{c}$ due to
effects of localization. The distance from the \lq\lq critical\rq\rq\/ point can be
characterized by the parameter

\begin{equation}
\delta =\frac{V_{g}-V_{g}^{c}}{eV_{g}^{c}}\sqrt{V_{g}^{c}\frac{e^2}
{2\epsilon_{ox}d}}
\propto
\frac{n-n_c}{n_c},
\end{equation}
where $n$ is the electron concentration.
$\rd$ in this model can be written as
\beqa
\rho _{{\rm d}}=\rho _{1}+{\bar\rho}\left( T/T_{0}\right) ^{q}\times \left\{ 
\begin{array}{ll}
\left( 1+|\delta |/cT\right) ^{q},\;{\rm for}\;\delta \leq 0; &  \\ 
e^{-\delta /T},\;{\rm for}\;\delta >0, & 
\end{array}
\right. 
\label{rdtrap}
\eeqa
where $q=1/2$, $c\approx 0.62$, and ${\bar\rho}$ depends on the
microscopic details of the model (trap concentration, etc.).
$\rho _{1\text{ }}$ describes the T-independent part of disorder, arising
from, e.g., residual impurities and interface roughness.
When $\mu$
is above the trap level ($\delta >0$,\lq\lq metallic\rq\rq\/ regime)
the number of charged traps $N_+$, and thus $%
\rho _{{\rm d}}$ decreases with $T$ exponentially. When the
chemical potential is below the trap level
($\delta <0$,\lq\lq insulating\rq\rq\/ regime), 
$N_+$ and $\rd$ saturate for $T\ll |\delta |$.
In the \lq\lq critical\rq\rq\/ region ($|\delta |\leq T$), $\rho _{{\rm %
d}}\propto T^{q}$. 

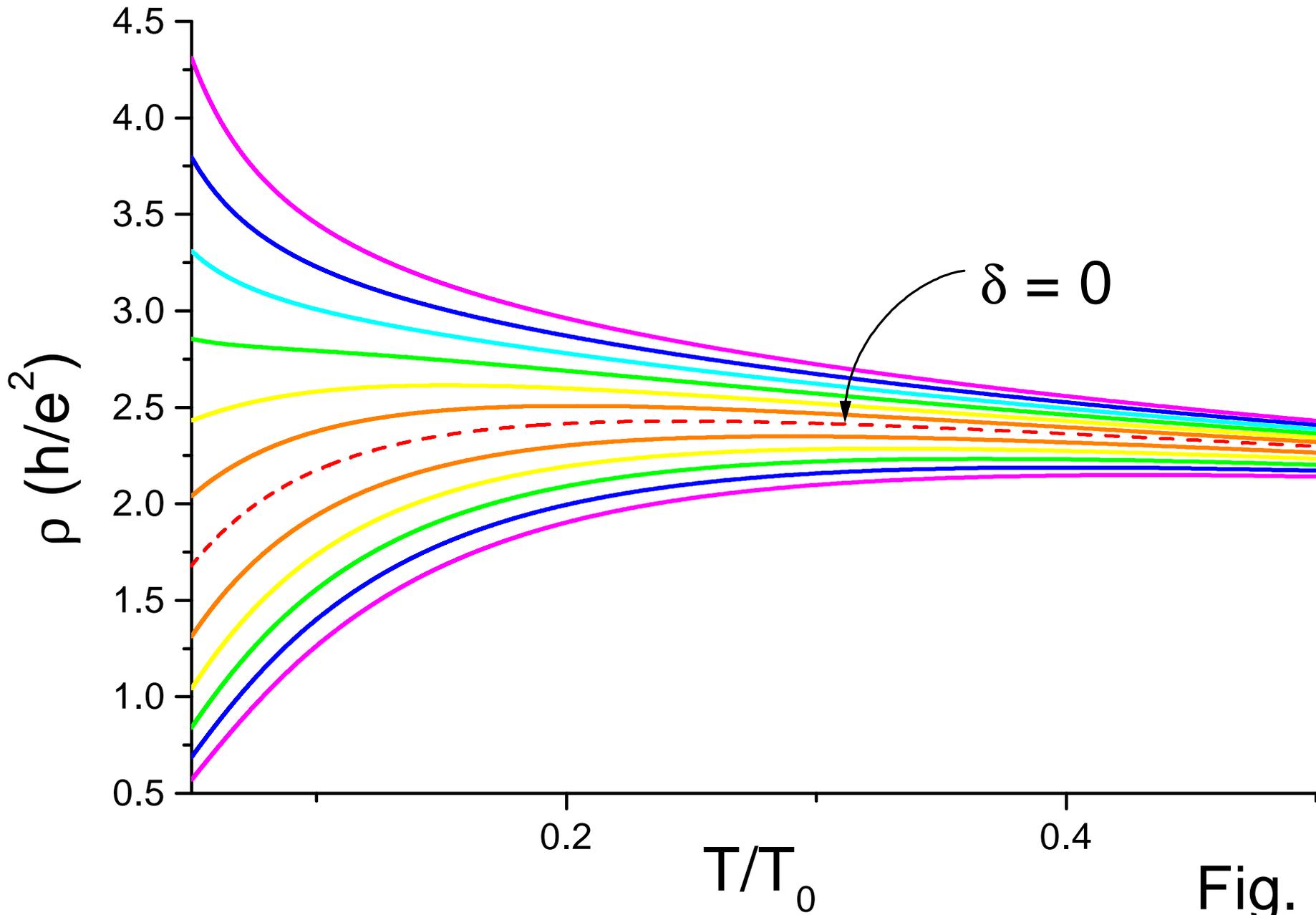
\begin{figure}
\caption{
Illustration of an MIT in the model of Ref.~[10].
$p=3,q=1/2,\gamma=0.5,{\bar\rho}=2.0,\rho_1=0.01$. From top to bottom: 
}
\label{fig:mit1}
\end{figure}

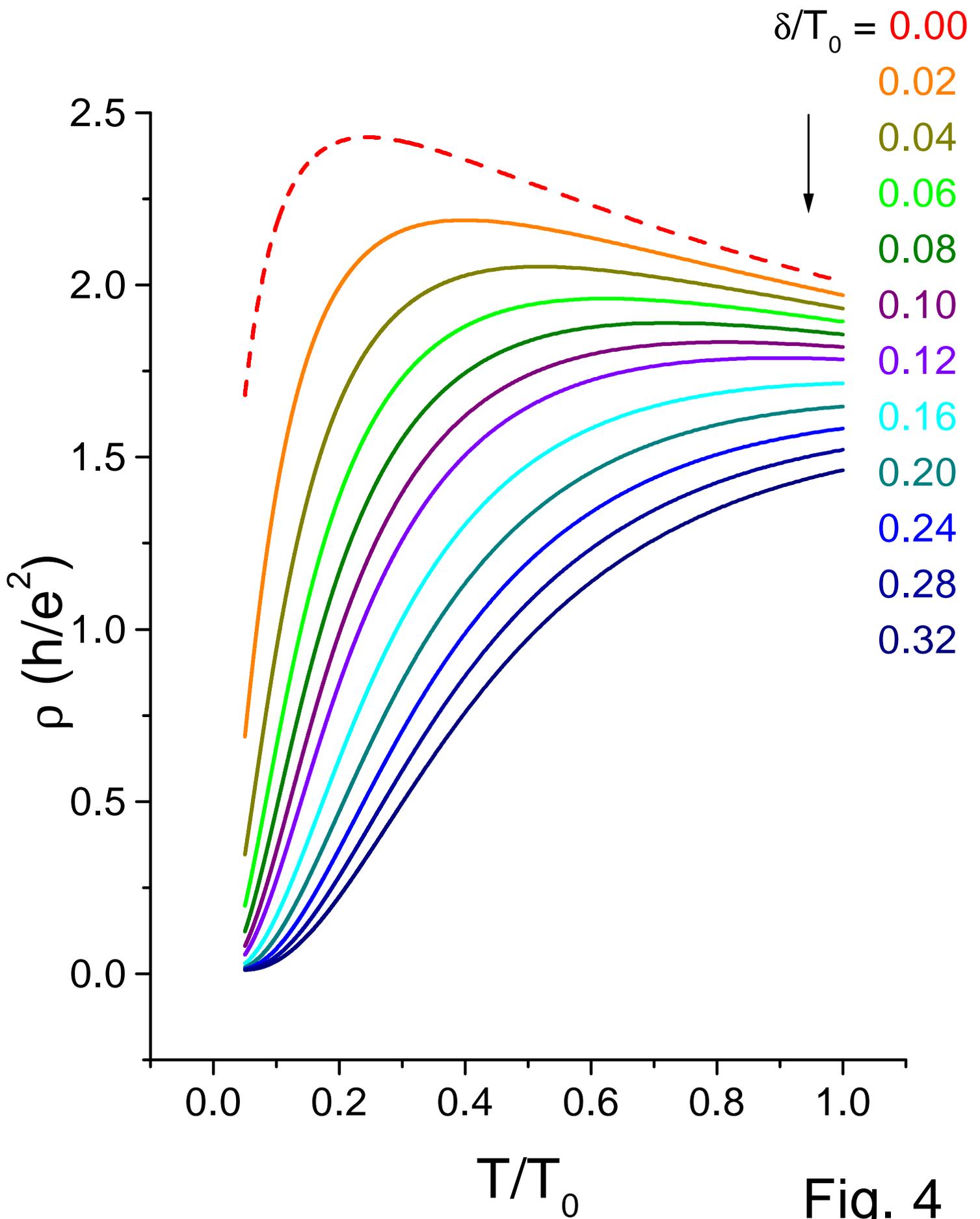
\begin{figure}\caption{Evolution of the maximum in $\rho(T)$
for the same parameters as in Fig.~\ref{fig:mit1}. }
\label{fig:max}\end{figure}

An illustration of an MIT-like behavior, corresponding 
to $\rd$ given by Eq.~(\ref{rdtrap}),
is presented in Fig.~\ref{fig:mit1}
for ${\bar{\rho}}=2.0$ and $p=3$.
For this choice of parameters, localization prevails over the
metallic-like temperature dependence over a range of \lq\lq metallic\rq\rq\/
densities ($\delta>0$) and $\rho (T)$
exhibit a maximum at $T=0.1..0.2T_{0}$ (Fig.~\ref{fig:max}).  As $%
\delta $ becomes negative and increases in its absolute value (which
corresponds to increasing $\rd$), the maximum turns over into a
monotonic, insulating-like $T$-dependence of $\rho $.

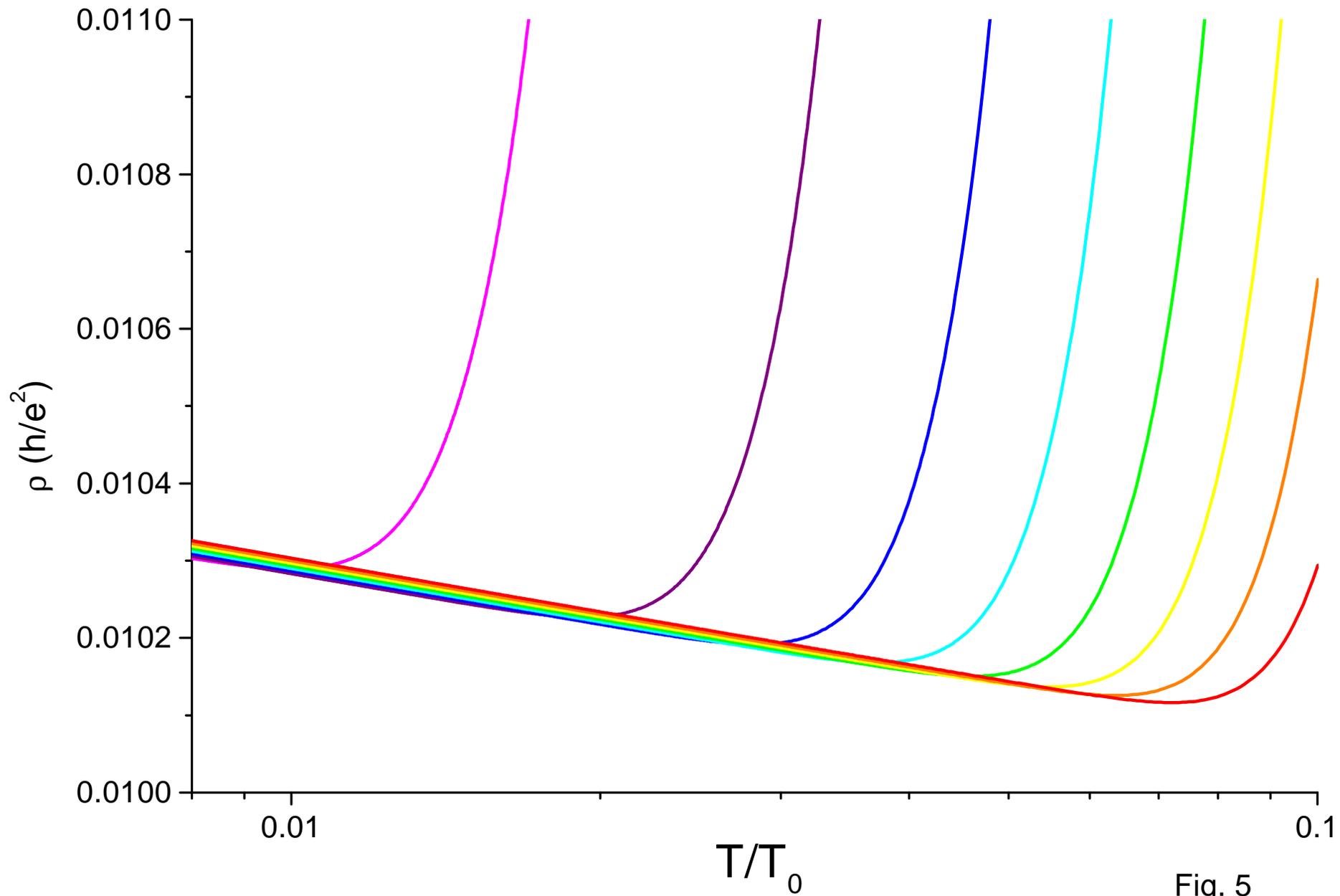
\begin{figure}
\caption{Evolution of the WL minimum in
$\rho$ for the same parameters as in
Fig.~\ref{fig:mit1}.
From left to right: $\delta/T_0=0.1\dots 0.8$ with increment $0.1$.}
\label{fig:min}\end{figure}

The $\rho(T)$ behavior at relatively high temperatures
$T=(0.1-1)T_0$ leaves one with
an impression of a true MIT. As $\rho_d(0)\neq 0$, however,
curves which look \lq\lq metallic\rq\rq
\/at higher $T$, exhibit localization behavior at lower $T$.
Fig.~\ref{fig:min} shows the behavior of  \lq\lq metallic\rq\rq\/ curves
in a wide temperature range. As was explained earlier, the
temperature of the minimum is higher for more \lq\lq metallic\rq\rq\/
 curves. In the
experiment, the high-density, low-resistance curves show an upturn at low
temperatures, whereas no such upturn is observed for lower-density,
higher-resistance curves. Fig.~\ref{fig:min} suggests a simple, if not trivial,
explanation of this effect: if the lowest temperature attained experimentally
is still higher than $T_{\text{min}}$, the curve may be interpreted as
\lq\lq metallic\rq\rq\/, whereas in fact it is \lq\lq insulating\rq\rq\/.

The work at
Princeton University was supported by ARO MURI DAAG55-98-1-0270. D.\ L.\ M.
acknowledges the financial support from NSF DMR-970338. V.\ M.\ P.\
acknowledges the support by RFBR via Program \lq\lq\ Physics of solid-state nanostructures\rq\rq\/,
INTAS, and NWO.

\newpage 
\begin{figure}\caption{
$\rho(T)$ (solid) corresponding to the
Drude resistance $\rho_{{\rm d}}(T)$ [Eq.~(\ref{power})] (dashed) for
${\bar\rho}=1,2,3$. $p=3,q=1/2,\gamma=0.5$. For this value of $p/q$, 
the critical value
of ${\bar\rho}$ for the existence of the maximum $\rho_c\approx
1 $. Inset: characteristic resistance $\rho_{c}$ as a function of 
$p/q$.
}
\label{fig:powera}\end{figure}

\begin{figure}
\caption{
Variation of the resistivity minimum with
parameter ${\bar\rho}$ in Eq.~(\ref{power}). $\rho%
_1=0.1 $. From top to bottom: ${\bar\rho}=0.5,1,2$.
}
\label{fig:minpower}\end{figure}

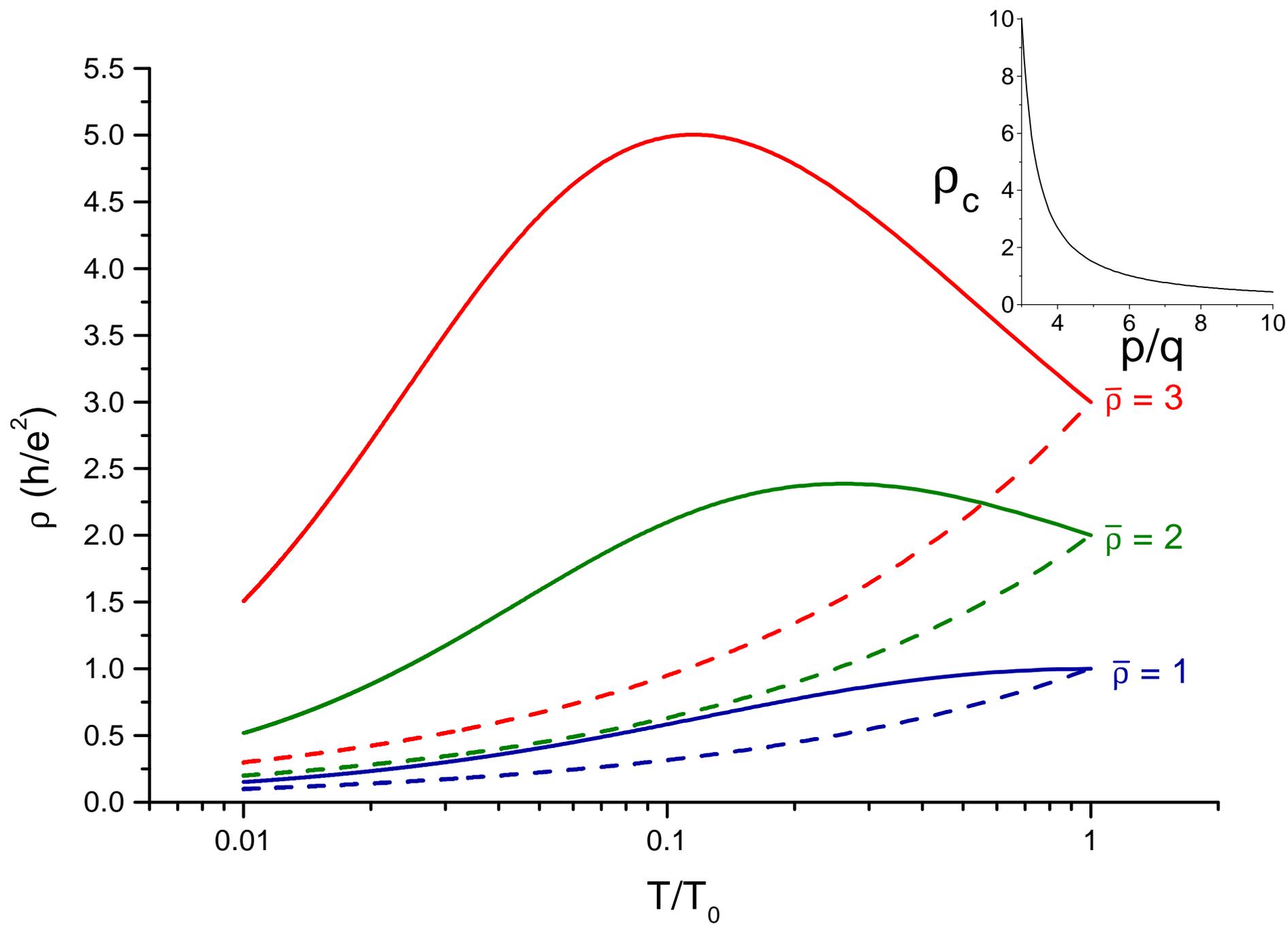
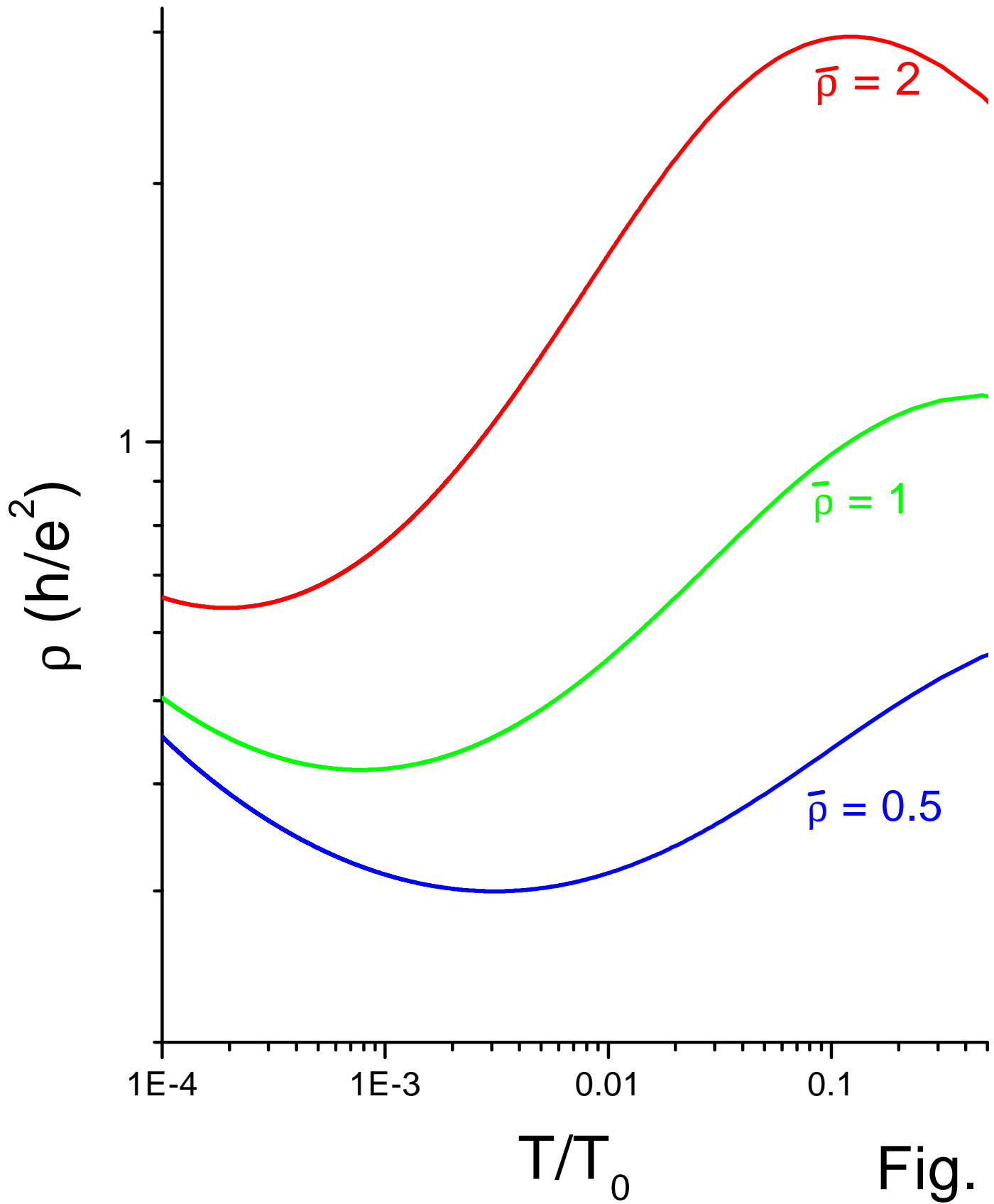
\end{document}